\documentclass[prb,twocolumn,aps,showpacs]{revtex4-1}
\usepackage{graphicx}
\usepackage{amsmath}
\usepackage{amsfonts}
\usepackage{ulem}
\usepackage{array}

\newcommand{\be}{\begin{equation}}
\newcommand{\ee}{\end{equation}}
\newcommand{\bea}{\begin{eqnarray}}
\newcommand{\eea}{\end{eqnarray}}

\newcommand{\comment}[1][1]{}

\begin{document}
\title{A Scaling Hypothesis for Modulated Systems}
\author{O. Portmann$^1$, A. G\"olzer$^1$, N. Saratz$^1$, O. Billoni$^{1,2}$, D. Pescia$^1$, and A. Vindigni$^1$}
\affiliation{Laboratorium f\"ur Festk\"orperphysik, ETH Z\"urich, 8093 Z\"urich, Switzerland\\
$^2$ Facultad de Matem\'atica, Astronom\'{\i}a y F\'{\i}sica (IFEG-CONICET), 
Universidad Nacional de C\'ordoba,\\ Ciudad Universitaria, 5000 C\'ordoba, Argentina}
\date{\today}
\begin{abstract}
We propose a scaling hypothesis for pattern-forming systems in which modulation of the order parameter results from 
the competition between a short-ranged interaction and a long-ranged interaction decaying with some power $\alpha$ of 
the inverse distance.  With $L$ being a spatial length characterizing the modulated phase, all thermodynamic quantities 
are predicted to scale like some power of $L$, $L^{\triangle (\alpha\!,d)}$. 
The scaling dimensions $\triangle (\alpha\!,d)$ only depend on the dimensionality of the system $d$ and the exponent $\alpha$. 
Scaling predictions are in agreement with experiments on ultra-thin ferromagnetic films and computational results. 
Finally, our scaling hypothesis implies that, for some range of values $\alpha\!>\!d$, Inverse-Symmetry-Breaking transitions 
may appear systematically in the considered class of frustrated systems.  
\end{abstract} 

\maketitle

\section{Introduction}
The emergence of modulated phases is a general motive in chemistry, biology, and physics~\cite{Seul,Muratov}. 
The modulated order parameter may represent quantities as diverse as the spin density~\cite{Garel,Poki,Van}, the 
charge density in any type of strongly correlated classical or quantum system~\cite{Muratov,Viot,Nussinov,Kivelson}, 
the volume fraction of diblock copolymers, the concentration of amphiphilic molecules and other chemical species~\cite{Muratov,Connell}, 
or dipolar bosons in an optical lattice~\cite{Menotti}. However, modulated systems tend to show common characteristics 
such as the morphology of the various patterns and the occurrence of transitions among them \cite{Seul}, as in liquid 
crystals~\cite{Harrison} or two-dimensional melting phenomena~\cite{KTNHY}. This tendency to common behavior and the 
scaling properties discovered recently in experiments on ferromagnetic ultra-thin films~\cite{Niculin} with modulated magnetization indicate that ``universal'' underlying principles possibly exist~\cite{Nussinov,Kivelson,Lieb}.\\ 
Here we consider a system embedded in a $D$-dimensional space described by an order parameter which is a function of $d\!\le\!D$ spatial variables. We refer to a situation in which a modulation of the order parameter takes place due to the competition between interactions of rather general types acting at different spatial scales. 
Domains carrying alternating signs of the order parameter thus appear. Calling $L$ the linear size of such domains, 
other possible characteristic spatial scales of the domain pattern are assumed to be proportional to $L$ itself (see Fig.~1).   
Alternatively, any other length scale characterizing a domain pattern can be defined as $L$. 
To be concrete, a typical pattern associated with a modulated order parameter can be thought as made of, e.g., stripes in
thin films or lamellae in bulk materials characterized by the order parameter alternating from one sign to the other when
moving along one direction in space. $L$ is in this case the width of the stripes or the thickness of lamellae. Other
possible arrangements are bubbles or cylinders carrying one sign of the order parameter embedded into a background of
opposite sign~\cite{Muratov}. In this case, $L$ shall be the distance between the centers of domains and the radius of each domain is assumed to be proportional to $L$ itself. 
Other patterns are possible -- some of them irregular~\cite{Muratov} -- but in the present paper all of them shall be 
the result of the competition between a short-ranged interaction, favoring a uniform order parameter, 
and a weak, but long-ranged, frustrating interaction. We assume the long-ranged interaction to decay with some power 
$\alpha$ of the inverse distance. An example of such interactions is the Coulomb interaction, with $\alpha\!=\!1$,
generating the Coulomb Frustrated Ising Ferromagnet (CFIF)~\cite{Viot}. Similarly, the dipolar interaction between spins,
originating from magnetostatics, may produce the Dipolar Frustrated Ising Ferromagnet (DFIF)~\cite{Ale}. For instance, thin films magnetized perpendicularly to the plane can be represented  with this model and an exponent $\alpha \!=\!3$.   \\
The outline of the paper is the following: 
In Section~\ref{sec_II}, we introduce the model Hamiltonian for general $(\alpha\!,d)$ and a coarse-grained version of it of 
the Landau-Ginzburg-Wilson type. 
In Section~\ref{sec_III}, we formulate a scaling hypothesis, which essentially consists in postulating that the spatial
 profile of the order parameter is invariant 
under rescaling of the characteristic length $L$. 
In Section~\ref{sec_IV}, we discuss the consequences of this scale-invariance hypothesis and show how the dependence on $L$ propagates to all the 
physical quantities with appropriate scaling exponents; such exponents only depend on $d$ and $\alpha$.  
When applied to the concrete experimental situation of ferromagnetic ultra-thin films, our scaling results agree with experimental findings~\cite{Niculin,Oliver1,Oliver2}.  
In Section~\ref{sec_V}, we show how the Coulomb and the dipolar interaction in a slab reduce asymptotically to a form of the type 
considered here.   
In Section~\ref{sec_VI}, we provide some numerical confirmations of the validity of our scaling analysis based on mean-field calculations and Monte-Carlo simulations.  
In Section~\ref{sec_VII}, we discuss the stability of our results against a perturbing displacement field which, for instance, destroys positional order by means of the Landau-Peierls instability.  
Finally, in Section~\ref{sec_VIII}, we propose the existence of a range of values $(\alpha\!,d)$ for which anomalous 
re-entrance phenomena involving Inverse Symmetry Breaking~\cite{Schupper,Weinberg} may occur systematically. 
The main features of such processes, as they emerge from our model, are also discussed in comparison with conventional Symmetry-Breaking mechanisms~\cite{Weinberg,Landau,Nambu,Rivers,Wilson}. 
\begin{figure}
\includegraphics{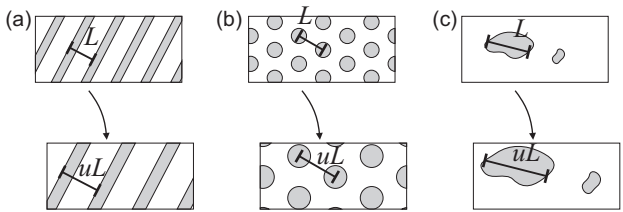}
\caption{Sketch of different patterns with a characteristic spatial scale $L$. The arrows indicate the process of rescaling them by a factor $u$. 
\label{figure1_cartoon}}
\end{figure}
\section{The effective Hamiltonian\label{sec_II}}
What we have in mind is a cubic lattice with lattice constant $a$ embedded in a $D$-dimensional space. The lattice
 sites are occupied by Ising variables $\sigma=\pm 1$, coupled by a short-ranged interaction of strength $J\!>\!0$, which
 favors the same value of the variable for nearest neighboring sites. Considered alone, the coupling $J$ establishes the
 well-known Ising model on the lattice. In addition, we introduce a weaker but long-ranged interaction 
(of strength $\lambda$) which favors opposite values of the Ising variables between any two sites, including distant ones.
 The long-ranged interaction is thus in competition with the short-ranged (exchange) interaction. The strength of the 
long-range interaction shall decay as some power $\alpha$ of the inverse distance between two sites. Finally, an external,
 uniform field couples linearly with the Ising variables. The external field is also in competition with the long-ranged
interaction as it favors a spatially uniform state. Experimentally, the external field can be easily tuned -- in a spin
 system for instance -- against the long-ranged interaction to produce transitions  between different patterns until, at
 some critical field, the uniform state is reached~\cite{Niculin}. The Hamiltonian describing this system reads:
\begin{eqnarray}
\label{discrete_lattice_Ham}
{\cal H} &=& -\frac{J}{2}\sum_{\langle i,j\rangle}\sigma_i\, \sigma_j + \frac{\lambda}{2}\sum_{i\ne j}\frac{\sigma_i \,\sigma_j}{| r_{ij}|^\alpha} - h\sum_i \sigma_i,
\label{Ham}
\end{eqnarray} 
where $\langle i,j\rangle$ means that the sum involves only nearest neighbors. The remaining sum extends over all lattice
 sites in the $d$-dimensional system. The Hamiltonian~\eqref{discrete_lattice_Ham} is suitable for discrete-lattice
 calculations, e.g. mean-field calculations~\cite{Ale} or Monte-Carlo simulations~\cite{Cinti} (see Section~\ref{sec_VI}). 
The scaling hypothesis we would like to introduce is more transparent when a coarse-grained, Landau-Ginzburg-Wilson (LGW)
 version of the lattice Hamiltonian in Eq.~\eqref{Ham} is used. Let the lattice extend over macroscopic lengths $\Lambda$
 along $d\leq D$ directions; along the remaining $(D\!-\!d)$ directions the system has a finite thickness $\delta\ll\Lambda$
 and the scalar field takes uniform values. The LGW-Hamiltonian proposed in Ref.~\onlinecite{Muratov} reads
\begin{eqnarray}
\mathcal L [\sigma(\vec x),\sigma_0(T)]&=&\frac{J}{2}\,\delta^{D-d}\sigma_0^2(T)\int(\nabla\sigma(\vec x))^2 d^d\!x \nonumber\\  
&-&zJ\delta^{D-d}\,\sigma_0^4(T)\int f[\sigma(\vec x)]d^d\!x \nonumber\\ 
&+&\!\frac{\lambda}{2}\delta^{2(D-d)}\, \sigma_0^2(T)\!\!\iint\!\frac{\sigma (\vec x)\,\sigma (\vec x')}{|\vec x\!-\! \vec x'|^\alpha }\,d^d\!x\,d^d\!x'\nonumber\\
&-&\!h\,\delta^{D-d}\,\sigma_0(T)\int \sigma (\vec x) d^d\!x,   
\label{functional}
\end{eqnarray}   
where $\vec x\doteq(x_1,...,x_d)$ is a vector in the $d$-dimensional space, $z$ is the number of nearest neighbors and 
\begin{eqnarray}
f[\sigma(\vec x)]\!=\!\frac{\sigma^2\!(\vec x)}{6}\!-\!\frac{\sigma^4\!(\vec x)}{12}\,.
\end{eqnarray}   

All spatial lengths considered here are given in units of the lattice constant $a$. The gradient term in the first line 
mimics the short-ranged interaction; at low temperature, when domain walls are sharp, it is inaccurate and a  
more suitable expression exists~\cite{Poki}. 
When one let the long-ranged interaction vanish (i.e. $\lambda\!=\!h\!=\!0$), for $d\!>\!1$ $\mathcal L$ produces a 
macroscopic phase separation~\cite{Muratov} with a spatially uniform order parameter appearing below a critical finite temperature 
$T_C$: $\sigma_0(T)\!\neq\! 0$ and $\sigma(\vec x)\!=\!+1$ or $\sigma(\vec x)\!=\!-1$, $\forall \vec x$.  When $\lambda\!\not=\!0$, 
we expect a spatially non-uniform distribution of the order parameter (so called micro-phase separation, in opposition to macroscopic phase separation~\cite{Kivelson}):  
modulated patterns thus form with one, or more, characteristic length scales.   
In this paper, we have in mind the situation where $J\!\gg \! \lambda$, corresponding to concrete experimental 
examples~\cite{Poki,Niculin,Ale,cc}, that produces typical characteristic lengths much larger than the lattice constant.  
Notice, however, that the results we are going to derive also apply to Monte-Carlo simulations performed on a lattice for  
ratios $3\!<\!J/\lambda\!<\!5$ (see Section~\ref{sec_VI}).\\ 
Modulated phases can also arise from competing \textit{short-ranged} interactions, as 
for the ANNNI model. Such modulated phases may share some features with the Frenkel-Kontorova model~\cite{Frenkel-Kontorova}, i.e., 
commensurate, incommensurate and devil-staircase phases may appear~\cite{Bak}.   
By varying the temperature it is possible to induce transitions between different modulated structures~\cite{Selke}. 
For specific lattice geometries and ratios of the strength of competing interactions, the ANNNI model exhibits a so-called \textit{disorder point}~\cite{Stephenson}: 
below a certain temperature $T_D$ two-spin correlations decay monotonically, while for $T\!>\!T_D$ they decay in an  
oscillatory manner. The period of such an oscillation may depend on the temperature (disorder point of the 
first kind) -- a feature which appears also in our model. 
A similar change of behavior in spin correlations can be triggered by 
an external field in non-collinear anisotropic spin chains~\cite{Vindigni}.  
The models listed above are different from the one we consider here because there the competition 
occurs between \textit{short-ranged} interactions of different sign or between a short-ranged interaction and the applied field. 
Instead, in Eqs.~\eqref{discrete_lattice_Ham} and \eqref{functional} the competing interactions act on entirely different spatial scales.  \\
The kernel $G_\alpha\doteq|\vec x\!-\!\vec x'|^{-\alpha}$ appearing in the coarse grained functional must be considered as 
a distribution defined over a set of test functions behaving properly for $(|\vec x|,|\vec x'|)\rightarrow \infty$. 
It is locally integrable if $\alpha\!<\!d$. For $\alpha\!\geq\!d$ the distribution $G_\alpha$ has a singular behavior at 
$\vec x=\vec x'$. In Appendix~\ref{Appendix_A}, we show how this distribution can be analytically continued to any 
real $\alpha$ and find, for the physically well-defined kernel, the Fourier transform  
\begin{equation}
G_\alpha=\frac{1}{| x-x'|^\alpha} =\int  G_\alpha(\vec k) \,e^{-i \vec k\cdot (\vec x-\vec x')} \,d^dk \,.
\label{kkernel}
\end{equation}
We will show that $G_\alpha(\vec k)$, up to logarithmic corrections for special values of $\alpha$ (see Appendix~\ref{Appendix_A}), is an \textit{homogeneous} distribution, i.e., for any $u$ real 
\begin{equation}
G_\alpha(u\,\vec k) = u^{\alpha-d}\, G_\alpha(\vec k)\,.
\label{inv}
\end{equation}
\section{The scaling hypothesis\label{sec_III}}
We now assume that\begin{itemize}
\item\label{assume_1} 
the scalar field -- $\sigma(\vec x,L)$ henceforth -- is non-uniform and characterized by a typical length scale $L$ 
\item\label{assume_2} $\sigma(\vec x,L)$ is invariant with respect to a rescaling of all lengths 
by some constant $u$: 
\begin{equation}
\sigma (u\, \vec x,u\, L) = \sigma(\vec x,L) \,.
\label{scaling}
\end{equation}
\end{itemize} 
These assumptions effectively restrict the microscopic configurations which can be expressed through the scalar field 
$ \sigma(\vec x)$; thus, the actual underlying hypothesis is that spin profiles fulfilling 
the property in Eq.~\eqref{scaling} represent the \textit{physically relevant} configurations.   
Note that the ground-state spin configuration of the Hamiltonian~\eqref{discrete_lattice_Ham} 
falls in the class of functions described by Eq.~\eqref{scaling}, as proved rigorously for $\alpha\!>\!d$ and $d\!=\!1$ in Ref.~\onlinecite{Lieb,Giuliani}.    
The same authors suggest that this property is probably true for any $d$~\cite{Giuliani_Granada}.   
The latter statement is supported by analytic calculations in which selective highly symmetric configurations are compared 
and by several numerical results~\cite{Viot,Van}. 
Our aim is to discuss some consequences of the scaling hypothesis defined by Eq.~\eqref{scaling} and compare them with 
experiments on magnetic films and numerical results. 
In Fig.~1  we sketch some examples of what we mean by $L$: 
it can be the period of modulation in a stripe-like pattern (Fig.~\ref{figure1_cartoon}a) or  
the linear size of the unit cell for any periodic pattern (see bubbles in Fig.~\ref{figure1_cartoon}b) or  
the characteristic size of a ``black'' droplet in a white background (Fig.~\ref{figure1_cartoon}c), etc.  
The effect on $\sigma(\vec x,L)$ of rescaling all lengths by a constant $u$ is also sketched in Fig.~1. 
As a consequence of the scaling hypothesis in Eq.~\eqref{scaling}, the scalar field $\sigma(\vec x,L)$ is in fact a function of $\vec x/L$ only. By substituting the variable $\vec y \!=\! \vec x/L$  
in Eq.~\eqref{functional}, the Hamiltonian (per unit volume $V_d\doteq\delta^{D-d}\,\Lambda^d$)  assumes the scaling form 
\begin{eqnarray}
\ell [\sigma(\vec y),L,T,h] &=& J L^{\triangle_J}\,\sigma_0^2(T)\, {\ell}_J[\sigma(\vec y)]\nonumber\\
&-& z J L^{\triangle_f}\,\sigma_0^4(T)\,{\ell }_f[\sigma(\vec y)] \nonumber\\
&+& g L^{\triangle_g }\,\sigma_0^2(T)\,{\ell }_g[\sigma(\vec y)]\nonumber\\
&-& h L^{\triangle_h}\,\sigma_0(T)\,{\ell }_h[\sigma(\vec y)]
\label{functionals}
\end{eqnarray}
where  $g \!= \!\delta^{D-d}\lambda$. 
The four functionals read  
\begin{eqnarray}
{\ell }_J[\sigma(\vec y)]&=& 
\left(\frac{L}{\Lambda}\right)^d\int_{-\frac{\Lambda }{2L}}^{\frac{\Lambda}{2L}} (\nabla \sigma(\vec y))^2 d^dy\nonumber\\  
{\ell }_f[\sigma(\vec y)]&=& 
\left(\frac{L}{\Lambda}\right)^d\int_{-\frac{\Lambda }{2L}}^{\frac{\Lambda}{2L}} f[\sigma(\vec y)] d^dy\nonumber\\
{\ell }_g[\sigma(\vec y,\vec y')]&=&
\left(\frac{L}{\Lambda}\right)^d\iint_{-\frac{\Lambda }{2L}}^{\frac{\Lambda }{2L}} \sigma (\vec y)\sigma (\vec y') d^dy d^dy' \nonumber\\
&\times&\int   G_\alpha(\vec k)\, e^{-i \vec k \cdot(\vec y-\vec y')}\,d^dk\nonumber\\
{\ell }_h[\sigma(\vec y)]&=&
\left(\frac{L}{\Lambda}\right)^d\int_{-\frac{\Lambda }{2L}}^{\frac{\Lambda }{2L}}  \sigma(\vec y) d^dy
\label{sfunctional} 
\end{eqnarray}
and are independent of $L$ in the thermodynamic limit $\Lambda\!\rightarrow \!\infty $, leaving the $L$-dependence of $\ell [\sigma(\vec y),L]$ solely to the pre-factors of the type $L^{\triangle_{\sharp}}$, 
with a scaling dimension $\triangle_{\sharp}$ specific to each interaction. 
The scaling dimensions can be computed explicitly:
\begin{equation}
\triangle_J\!=\!-\!1\,(-2)\quad   \triangle_g\!=\!d\!-\!\alpha\quad\triangle_f=\triangle_h\!=\!0\,.
\label{dim}
\end{equation}
The value of $\triangle_J=-1(-2)$ refers to sharp (extended) walls, which are expected to be realized at low (high) temperatures. \\
Note that the scaling form of Eq.~\eqref{functionals} is a consequence of $G_\alpha(\vec k)$ being homogeneous, Eq.~\eqref{inv}, which is true for most values of $\alpha$. 
In some special cases, e.g. in the one-dimensional lattice with Coulomb interaction, logarithmic corrections appear and must be considered \textit{ad hoc}.     

\section{Scaling results\label{sec_IV}}
One possible way of obtaining results from the scaling functional in Eq.~\eqref{functionals} is using a variational approach. The variational equation 
\begin{equation}
\frac{\delta\ell [\sigma(\vec y),L]}{\delta \sigma(\vec y)}=0 
\label{var_eq_1}
\end{equation}
produces the equilibrium profile $\bar\sigma_{L,T,h}(\vec y)$, which, in virtue of our scaling hypothesis, Eq.~\eqref{scaling}, 
should be almost independent of $L$. 
The dependence of the solution, $\bar\sigma$, on $(L,T,h)$ is only \textit{parametric} and results from the dependence of the functional $\ell$ on such parameters. 
Inserting $\bar\sigma_{L,T,h}(\vec y)$ into the Hamiltonian Eq.~\eqref{functionals} leads to a Landau free-energy functional ${\ell}(L,T,h)$. 
If $\bar\sigma_{L,T,h}(\vec y)$ is \textit{almost} independent of $L$, 
the dependence on $L$ in ${\ell }(L,T,h)$ is \textit{essentially} restricted to the 
pre-factors with scaling dimensions given by Eq.~\eqref{dim} and therefore scaling results can be immediately obtained. 
Typically, the solutions of the variational Eq.~\eqref{var_eq_1} are highly symmetric spin profiles which -- in general -- are not a realistic representation of 
experimental patterns. However, in Section~\ref{sec_VII} we provide arguments that our scaling results are not affected by fluctuations.  
Moreover, both experiments~\cite{Niculin} and Monte-Carlo simulations presented in Section~\ref{sec_VI} confirm those results for accesible observables related to $L$ and selected values of $\alpha$ and $d$.   
In the following, we list some scaling results. 
 
\noindent {\bf 1.} 
We define the average characteristic length, $\bar L(T)$, as the solution of the equation $\partial {\ell }(L,T,h)/\partial L\!=\!0$.  
From Eq.~\eqref{functionals},  $\bar L(T)$ scales as:
\begin{equation}
\bar L \sim \Bigl(\frac{J}{g}\Bigr)^{\Gamma_{\bar L}}\,,\quad\text{with}\quad \Gamma_{\bar L}=\frac{1}{d-\alpha -\triangle_J}\,.
\label{lambda}
\end{equation}
Note that, in general, walls are very thin at low temperatures where we expect $\triangle_J\!=\! -1$. At higher temperatures walls broaden and $\triangle_J$ crosses over to the value $-2$. This produces a cross-over of the exponent 
$\Gamma_{\bar L}$ when the temperature is raised. 
Thus, in the limiting case considered in this paper, $J\!\gg\!g$, the scaling Eq.~\eqref{lambda} predicts a decrease of the equilibrium modulation length with increasing temperature~\cite{Ale}. 
Particularly striking is the cross-over (and the corresponding decrease of $\bar L(T)$) for the two-dimensional thin 
DFIF model, which has effectively $\alpha\!=\!3$, $d\!=\!2$ and $\delta\!\ll\!\bar L$ (see Section~\ref{sec_V}). 
This is a well-studied model~\cite{Van,Singer} and has an experimental counterpart~\cite{Niculin,Qiu}. For $\triangle_J\!=\!- 1$ ($T\!\simeq\! 0$), we have $\Gamma_{\bar L }\!=\!\infty $, which suggests an exponential dependence on $J/g$ at low temperatures. 
Detailed calculations indeed confirm that the period of modulation of a pattern that minimizes the energy~\eqref{discrete_lattice_Ham}   
diverges exponentially with $J/g$~\cite{Poki,Kivelson}.  
This property holds true, e.g., for stripe, checkerboard~\cite{Van} and bubble~\cite{Niculin_thesis} configurations in line with the scaling result of Eq.~\eqref{lambda}. 
At higher temperatures, the scaling exponent of $\bar L$ approaches just one, showing that $\bar L$ scales \textit{linearly} with $J/g$~\cite{Ale}. For $J \!\gg\! g$, this implies that the modulation length decreases by several orders of magnitude with increasing temperature~\cite{Ale,Singer,Qiu,Niculin}.\\
Another important consequence of Eq.~\eqref{lambda} is that a positive scaling dimension $\Gamma_{\bar L}$ is required in order for the modulation length to be physically plausible, i.e. $\bar L\!>\!1$, 
in the limit $J\!\gg\!g$. This suggests a possible bound for the existence of a modulated state. Indeed, we show in Appendix~\ref{Appendix_B} that in the limit
$J\!\gg\!g$ a modulated state is only stable for values of $\alpha$ such that  
\begin{equation}
\alpha\leq d-\triangle_J \,.
\label{stability}
\end{equation}
In Ref.~\onlinecite{Lieb} a rigorous stability line for \textit{any} value of the ratio $J\!/\!g$ was derived for $d\!=\!1\!$ and $T\!=\!0$.  For $J\!\gg \!g$, this rigorous stability line converges toward the limit established by 
Eq.~\eqref{stability} for $\triangle_J\!=\!-1$ (scaling exponent at $T\!=\!0$). It is worth noting that Eq.~\eqref{stability} allows for a curious phenomenon:  
As $\triangle_J$ crosses over from $-1$ to $-2$ when the temperature is raised, a range of values of $\alpha$ exists for which a modulated phase might become stable at finite temperatures
while being forbidden in the ground state. 
Something analogous occurs in the ANNNI model where an antiferromagnetic long-range-ordered phase 
can intervene between the ferromagnetic and the paramagnetic phase when the temperature is increased for specific lattice geometries and strengths of the competing interactions~\cite{Schupper,Stephenson,Selke}.  \\
From the fact that $\triangle_h$ appearing in Eq.~\eqref{functionals} is zero one may be led to conclude that $\bar L$ is independent of $h$. 
In fact, this is not exactly true because $h$ affects $\bar \sigma(\vec y,L,T,h)$.  
For the particular case of $d\!=\!2$ and $\alpha\!=\!3$, ground-state analytic computations show that in a wide range of fields the characteristic length $\bar L$ is indeed only weakly dependent on $h$~\cite{Niculin,Niculin_thesis}. 
However, close to the transition to the uniform state, $\bar L$ develops a singularity, which is evidently not captured by Eq.~\eqref{lambda}.  
Experimental data obtained away from the singularity also confirm that $\bar L$ is almost independent of $h$~\cite{Niculin_TBP}.\\
    
\noindent {\bf 2.}
In the absence of magnetic fields, the modulated state realized at equilibrium has the important property that regions with positive and negative sign of the order parameter are perfectly balanced, so that the asymmetry 
\begin{equation}
\bar A(T,h)\doteq\left(\frac{1}{\Lambda}\right)^d\int_{-\Lambda/2}^{\Lambda/2}  \bar \sigma(\vec x,T,h,L)\,d^dx
\end{equation}
vanishes exactly at $h=0$. $\bar A(T,h)$ is essentially the total ``polarization'' of the modulated system (divided by the volume of the system and by $\sigma_0(T)$). 
Starting from a modulated phase in $h=0$, for general $d$ and $\alpha$, when a small positive field is switched on 
the average volume occupied by domains with $\bar \sigma(\vec x)>0$ will increase at the expense of the volume of domains with $\bar \sigma(\vec x)<0$. 
In small fields, this unbalance can be accomplished by favoring microscopic configurations in which the domain walls are slightly shifted without varying their number (proportional to $1/L$)
nor their shape. 
Formally, such changes only affect the field-dependent and the long-range-interaction terms of the functional~\eqref{functionals}. 
Consequently, the total change of the functional (at least of the $L$-dependent part) writes:  
\begin{eqnarray}
\delta {\ell}=\!\frac{1}{2}\!\left(\!g\bar L^{\triangle_g}\sigma_0^2(T) \frac{\partial^2 {\ell}_g}{\partial A^2}\Bigg|_{A=0}\!\right) \!A^2
\!-\! h\, \bar L^{\triangle_h}\sigma _0(T) \, A \,.
\end{eqnarray}
Minimizing with respect to $A$, we obtain 
\begin{eqnarray}
\bar A(h\sim 0,T))&\sim &\frac{h }{\sigma _0(T)\,\bar L(T,h)^{\triangle_g}} 
\label{A}
\end{eqnarray} 
or, for the corresponding susceptibility 
\begin{equation}
\chi_A(h=0,T)\doteq \frac{\partial \bar A}{\partial h}\Bigg| _{h=0}\sim  
\frac{1 }{\sigma _0(T)\,\bar L(T,h)^{\triangle_g}} \,,
\label{Chi}
\end{equation} 
Eqs.~\eqref{A} and~\eqref{Chi} contain a scaling prediction for measurable quantities.  
When $\bar A$ is plotted in a coordinate space $(\bar A,T,h)$ its values lie on a two-dimensional surface. However, according to Eq.~\eqref{A}, such values should collapse onto one single curve if $\bar A$ is plotted versus 
$h/(\sigma_0 \bar L^{\triangle_g})$. This data collapsing and the corresponding scaling exponent  $\triangle_g\!=\!d\!-\!\alpha$ 
have been verified experimentally in Ref.~\onlinecite{Niculin} for 
ultra-thin magnetic Fe films on Cu ($\alpha\!=\!3$, $d\!=\!2$ and $\delta\!\ll\!\bar L$). 
In Section~\ref{sec_VI}, we will present Monte-Carlo results that confirm the validity of the scaling relation of Eq.~\eqref{Chi} for $d\!=\!1$ and three different values of $\alpha$. \\

\begin{figure} 
\includegraphics{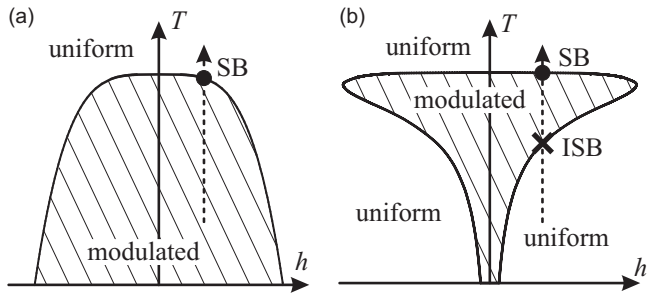}
\caption{Schematic phase diagrams in the $T-h$-plane for positive (a) and negative (b) values of the scaling exponent$\triangle$ as discussed in the text. In the hatched area a modulated phase is expected and the dashed arrows indicate possible paths referred to in the text. Circles mark regular symmetry breaking (SB) transitions, the cross in (b) marks inverse symmetry breaking (ISB).
\label{figure2_dome-funnel}}
\end{figure}
\noindent {\bf 3.}
At some threshold value, generally dependent on $T$, the external field (fourth term in Eq.~\eqref{functionals}) produces a transition between the modulated phase -- characterized by some $\bar L$ -- and   
the uniform state~\cite{Seul,Garel,Singer,Niculin}. In fact, the external-field energy, favoring a uniform state, is in competition with the long-range interaction, which favors modulated patterns. 
At zero field, modulated configurations have a lower energy and equilibrium patterns with $\bar A=0$ are realized. 
However, a finite field lowers the energy of other patterns with $\bar A\ne0$ or that of the uniform state ($\bar A = \pm 1$). 
Whether a modulated pattern or the uniform state is realized at equilibrium depends on the balance between the energy of the long-ranged interaction, which scales as 
$\sigma_0^2\,\bar L^{d-\alpha}$, and the energy of the external field, which scales as $\sigma_0\,\bar L^{0}$. 
Accordingly, we suggest that the transition field ($h_t$), at which a crossover from a modulated to the uniform phase occurs, should scale as 
\begin{equation}
h_t\sim \sigma_0(T)\,\bar L(T,h)^{\triangle_g-\triangle_h} \sim \sigma_0(T)\,\bar L(T,h)^{\triangle_g}\, .
\label{transition}
\end{equation}
This equation states that, while the phase transition lines in the $h\!-\!T$ plane might have some curvature (typically they have a dome-like shape~\cite{Singer}), they are straight lines 
when $h_t$ is plotted as a function of $\sigma_0 \,\bar L^{\triangle_g} $. 
This linear dependence and the corresponding scaling exponent $\triangle_g\!=\!d\!-\!\alpha$ have been verified experimentally in ultra-thin magnetic films~\cite{Niculin}.  
One remarkable feature of the system investigated in Ref.~\onlinecite{Niculin} is that $d\!=\!2$ and $\alpha\!=\!3$ so that the scaling dimension $\triangle_g$  
is \textit{negative}, leading to $h_t\propto \sigma _0/\bar L$. 
As in the magnetically ordered phase (away from the Curie temperature, $T_C$) $\bar L$ is typically decreasing much more strongly than $\sigma_0$ for increasing $T$, having $\triangle_g\!<\!0$  
implies that the transition lines in the $h\!-\!T$ plane have the shape of a ``funnel'' (see Fig.~\ref{figure2_dome-funnel}b) instead of resembling a dome (Fig.~\ref{figure2_dome-funnel}a).  
This funnel shape, predicted by our scaling hypothesis and verified experimentally~\cite{Niculin}, is anomalous with respect to the traditional phase diagrams of modulated systems  
and allows for Inverse Symmetry Breaking (ISB). 
Keeping in mind that the uniform phase has more symmetry elements than the modulated phase, a path in the $h\!-\!T$ plane through which the system passes from a modulated to the uniform phase with increasing $T$ 
corresponds to a Symmetry Breaking (SB) process (circle in Fig.~\ref{figure2_dome-funnel}a and ~\ref{figure2_dome-funnel}b). 
This is the ordinary scenario in thermodynamic phase transitions. Instead, when transition lines between a uniform and a modulated phase  
display a funnel shape in the $h\!-\!T$ plane, paths associated with ISB processes are also possible (cross in Fig.~\ref{figure2_dome-funnel}b). Some more comments on this important point will be given in Section~\ref{sec_VIII}.   \\

\noindent {\bf 4.}
An elastic constant used to characterize modulated matter is the compression modulus $B(T)$~\cite{Sornette,Poki} (also called Young modulus~\cite{Kleman}). 
It typically measures the energy cost associated with deviations from the equilibrium modulation length $\bar L(T)$ and is defined as
\begin{eqnarray}
B(T) \doteq \left. \bar L^2\frac{\partial^2 {\ell }(L,T)}{\partial L^2} \right |_{\bar L(T)} \,.
\label{modulus}
\end{eqnarray}
$B$ has also a scaling behavior, namely it has the same scaling dimension with respect to $\bar L$ as $h_t$: 
\begin{equation}
B\sim g\,\sigma _0(T)^2\, \bar L(T)^{\triangle_g} \,.
\label{scaling_B}
\end{equation}
\begin{figure}
\includegraphics{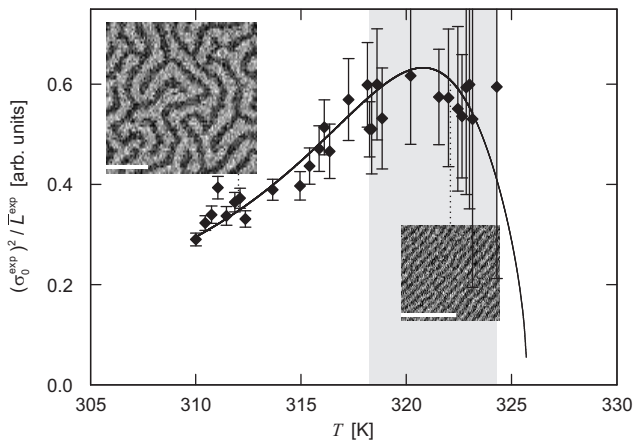}
\caption{$B^{exp}(T)$ obtained for an ultra-thin Fe film on Cu(001) by using $\sigma_0^{exp}$ and $\bar L^{exp}$ as determined from SEMPA (Scanning Electron Microscopy with Polarization Analysis) images in the scaling Eq.~\eqref{scaling_B}. The solid line results from individual fits to $\sigma_0^{exp}(T)$ and $\bar L^{exp}(T)$ respectively and is a guide to the eye. The temperature range with stripe domains is shaded. The two insets show representative SEMPA images of the labyrinthine and the stripe patterns. The length of the white bar is 10 $\mu$m, the thickness of the film 1.95 ML.}
\end{figure}
We are not aware of any direct measurement of $B$ on systems which can be modeled with the  Hamiltonian~\eqref{discrete_lattice_Ham}. 
However, \textit{assuming} the scaling Eq.~\eqref{scaling_B} to be true, the temperature dependence of $B$ can be 
deduced from the knowledge of $\sigma _0(T)$ and $\bar L(T)$. 
The latter quantities can be accessed experimentally, e.g., by analyzing the domain patterns in ultra-thin Fe films on Cu(001)~\cite{Oliver1} ($\alpha\!=\!3$ and $d\!=\!2$ giving $\triangle_g\!=\!-1$), so that 
it is possible to derive $B^{exp}=\left(\sigma_0^{exp}\right)^2 \left(\bar L^{exp}\right)^{-1}$ indirectly, see Fig.~3. 
Each data point corresponds to the value of $B^{exp}$ derived from one image. Within its error, the sequence of points clearly shows an enhancement of $B^{exp}$ with increasing temperature 
(in agreement with the mean-field calculation in Section~\ref{sec_VI}).  
From individual fits of the experimental $\sigma_0^{exp}(T)$ and $\bar L^{exp}(T)$ we can derive a smooth curve for $B^{exp}(T)$ in the same way, the solid line in Fig.~3. 
The decrease of $B(T)$ close to the experimental $T_C$ predicted in this way is not supported by individual data points and must therefore remain speculative at this stage. 
In the experiments a transition from a less symmetric labyrinthine pattern to a more symmetric stripe pattern (grey area in Fig.~3) is observed for increasing temperature. 
The fact that $B^{exp}$ also increases with increasing $T$ in the same region suggests the existence of some relationship between the two phenomena. 
A quantitative statement regarding such a relationship would require a deeper theoretical and experimental characterization of the specific system, 
which is beyond the scope of this paper. 
\section{Mapping of two physical interactions into  $G_\alpha$ \label{sec_V}} 
In this Section we show that the Coulomb and dipolar interactions take, asymptotically, the same form 
as $G_\alpha (\vec x -\vec x')$, Eq.~\eqref{kkernel}, when the order parameter $\sigma (\vec x)$ is a function of two variables only. 
The system we consider consists of a slab embedded in the real space ($D\!=\!3$), extending to macroscopic dimensions in the $(x_1,x_2)$-plane and with finite but variable thickness $\delta$ in the remaining direction. 
Along the latter direction, perpendicular to the plane of the slab, the scalar field $\sigma(\vec x)$ is assumed to be uniform; as a consequence,  
the problem is effectively two-dimensional ($d\!=\!2$). For this geometry, the dipolar (coupling constant $\Omega$) 
and Coulomb (coupling constant $Q$) interaction energies take the following forms for any thickness $\delta$: 
\begin{eqnarray}
G_\Omega(\vec \xi)\!&\!=\!&\!\frac{1}{\pi\,\delta^2}\int_0^{\infty }\cos(\vec k\cdot\vec \xi)\,\frac{1-e^{-\delta|\vec k|}}{|\vec k|}\,d^2k\\
G_Q(\vec \xi)\! &\!=\!&\!\frac{2}{\pi\,\delta^2}\!\int_0^{\infty}\cos(\vec k \cdot \vec \xi)\left(\frac{e^{-\delta|\vec k|} +\delta| \vec k|-1}{|\!\vec k|^3}\right) \, d^2k \nonumber
\label{Kernels_exact}
\end{eqnarray}
$\vec \xi\doteq \vec x-\vec x'$ being a vector in the plane of the slab. 
Here we label the physical quantities corresponding to the dipolar and Coulomb interactions with the relative coupling constants $\Omega$ and $Q$, rather than with
$\alpha$ (including the kernels $G$ of the interactions).   
Using the results of Appendix~\ref{Appendix_A}, we are able to evaluate some limiting cases:
\begin{equation}
\begin{split}  
&\frac{\delta}{L}\rightarrow \infty \quad G_\Omega^\infty(\vec k)\!\sim\!\frac{1}{\delta^2|\vec k|}\Leftrightarrow G_\Omega^\infty(\vec \xi)\sim \frac{1}{\delta^2 |\vec \xi|}\\
&\frac{\delta}{L}\rightarrow 0 \,   \quad G_\Omega^0(\vec k)\sim - |\vec k|\Leftrightarrow G_\Omega^0(\vec \xi)\sim \frac{1}{|\vec \xi|^3}\\
&\frac{\delta}{L}\rightarrow \infty \quad G_Q^\infty(\vec k)\sim \frac{1}{\delta |\vec k|^2}\Leftrightarrow G_Q^\infty (\vec \xi)\sim -\frac{1}{\delta} \ln(|\vec \xi|)\\
&\frac{\delta}{L}\rightarrow 0  \,  \quad G_Q^0(\vec k)\sim \frac{1}{|\vec k|}\Leftrightarrow G_Q^0(\vec \xi)\sim \frac{1}{|\vec \xi|} \,,
\label{Kernels_asympt}
\end{split}
\end{equation} 
where $L$ is the characteristic length introduced before. 
Two important statements are contained in Eq.~\eqref{Kernels_asympt}: \begin{enumerate}
\item in both limits, thin ($\delta/L\!\rightarrow\!0 $) and thick ($\delta/L\!\rightarrow\!\infty$) slabs, the kernels 
of the Coulomb and dipolar interactions behave like  $G_\alpha(\vec \xi)$ for large $|\vec \xi|$;  
\item the effective $\alpha$ can be varied by changing the thickness $\delta$. 
\end{enumerate}
The mapping stated above allows extrapolating the scaling results, deduced for the functional Eq.~\ref{functional}, to systems interacting via 
Coulomb or dipolar interaction within a slab geometry.    
In Table 1 we summarize some scaling results computed with the suitable $\alpha$ and for $d\!=\!2$ specific to this case. 
The appropriate values for $\alpha$ can be read out from the $\vec \xi$-dependence on the right-hand side of the equations. 
\begin{table}
\begin{center}
\begin{ruledtabular}
\begin{tabular}{lccccc}
&  & $\Gamma^{\triangle_J=-1}_{\bar L }$& $\Gamma^{\triangle_J=-2}_{\bar L }$&$\triangle_g$ \\
\phantom{\LARGE{A}} & $\alpha$ & $\frac{1}{3-\alpha }$ & $\frac{1}{4-\alpha }$ & $2-\alpha $  \\
 \hline
DFIF($\infty$) \phantom{\LARGE{A}}&1&$\frac{1}{2}$ &$\frac{1}{3}$& $\phantom{-}1$\\
DFIF ($0$) &$3$& $\infty$ & $1$ & $-1$ & \\
CFIF($\infty$) &$0$ &$\frac{1}{3}$ & $\frac{1}{4}$ & $\phantom{-}2$  \\
CFIF ($0$) &$1$& $\frac{1}{2}$  & $\frac{1}{3}$ &  $\phantom{-}1$ \\
\end{tabular}
\caption{
In the first column the effective $\alpha$ is given for the CFIF and DFIF at limiting thicknesses: ``0'' (``$\infty$'') stands for $\delta/L\!\rightarrow\!0 $ ($\delta/L\!\rightarrow\!\infty $). 
The corresponding scaling exponents $\Gamma_{\bar L}$ and $\triangle_g$ are given in the other columns.  
$\triangle_J\! =\!-1$ at $T\! =\!0$ and  $\triangle_J\! =\!-2$ at high $T$. 
\label{exponents}}
\end{ruledtabular}
\end{center}
\end{table}
Note that the thin ($\delta/L\rightarrow 0$) CFIF and the thick ($\delta/L\rightarrow \infty$) DFIF have the same effective $\alpha\!=\!1$.  
The scaling exponent $\Gamma_{\bar L}$ is given in Eq.~\eqref{lambda}, the two values of $\triangle_J$ corresponding to zero and high temperature respectively. 
$\triangle_g\!=\!d\!-\!\alpha$, appearing in Eqs.~\eqref{A}\eqref{Chi}\eqref{transition}\eqref{scaling_B},  is, instead, independent of $T$. 
As pointed out already in Section~\ref{sec_IV}, for the thin DFIF $\Gamma_{\bar L}\rightarrow \infty$ when $T$ approaches zero 
while $\Gamma_{\bar L}=1$ at high temperatures. This fact implies that $\bar L(T)$ varies over several orders of magnitude for realistic values of $J$ and $g$.  
\section{Numerical checks of the scaling hypothesis\label{sec_VI}}
In this Section we present some numerical checks of the scaling hypothesis that we performed for different values of $\alpha$ using the 
mean-field approximation, for a two-dimensional system, and the Monte-Carlo method for a spin chain.  
\begin{figure}
\includegraphics{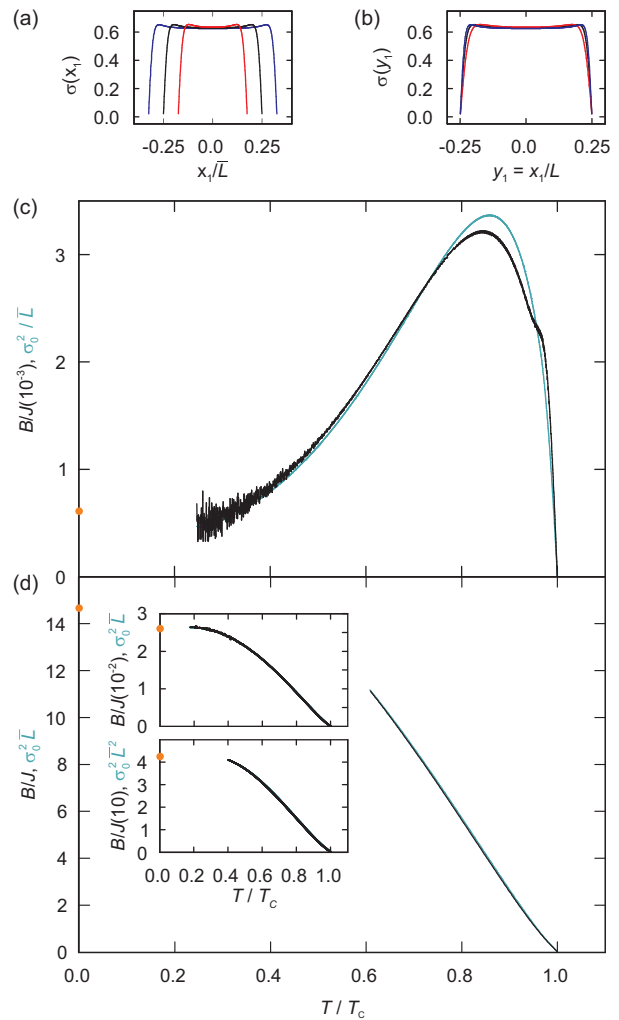}
\caption{Mean-field results (color online). (a) Profile of the order parameter in a thin DFIF ($\delta=1$, $\Omega/J=0.08$) for different values of $L$ ($0.7 \bar{L}$, $\bar{L}$ and $1.3 \bar{L}$). (b): Profiles from (a) as a function of $x_1/L$. (c): Numerical $B(T)$ for the thin DFIF (black curve, $\delta=1$, $\Omega/J=0.08$) together with the scaling result (light blue line). (d): Numerical and scaling curves $B(T)$ for the thick DFIF ($\delta=200$, $\Omega/J=0.08$, main plot) and the thin ($\delta=1$) and thick ($\delta=200$) CFIF ($Q/J=0.0001$, top and bottom insets, respectively). At low temperatures domain walls get sharper and sharper thus numerical results become unstable (see the text). 
Light orange circles on the vertical axis indicate the values of $B$ at $T=0$ obtained analytically. \label{figure1}}
\end{figure}
\subsection{Mean-field calculation for a 2d striped pattern} 
In the following, we present some mean-field results obtained for the systems introduced in Section~\ref{sec_V}.   
Thus in this paragraph we always deal with $d\!=\!2$. 
The mean-field approximation consists in solving the variational equation
\begin{equation} 
\frac{\delta{\mathcal L }[\sigma(\vec x),\sigma_0(T)]}{\delta(\sigma_0 \,\sigma(\vec x))} = 0\,;
\label{var_eq_2}
\end{equation}   
the solution $\sigma_0(T)\,\bar\sigma_{L,T,h}(\vec x)$ is then used to compute other physical quantities.  
In the present case, the functional to be minimized is the one in Eq.~\eqref{functional} with the exact kernels of Eqs.~\eqref{Kernels_exact} for the Coulomb and dipolar interactions 
in place of $1/|\vec x-\vec x'|^\alpha$. 
$\lambda$ is equal to $Q$ and $\Omega$ for the Coulomb and dipolar interactions, respectively (the definition of $Q$ and $\Omega$ is given in footnote~\cite{cc}). 
We used the Bragg-Williams form for the entropy in place of $f[\sigma(\vec x)]$ to obtain accurate results at low $T$. 
In the present calculation we set $h\!=\!0$. 
The variational problem defined by Eq.~\eqref{var_eq_2} differs from the one defined in  Eq.~\eqref{var_eq_1} 
because in the former the functional $\mathcal L$ is minimized with respect to the \textit{total} scalar field $\sigma_0\,\sigma(\vec x)$. 
We further restricted the configurations $\sigma(\vec x)$ to a mono-dimensional modulation along $x_1$, i.e., a striped pattern. 
We computed the order parameter, $\sigma_0$, and its profile, $\bar\sigma_{L,T,0}(x_1)$, which minimize the functional in Eq.~\eqref{functional} 
for a striped structure of arbitrary modulation period $L$  (solution of Eq.~\eqref{var_eq_2}).    
The equilibrium period of modulation, $\bar L$, can then be computed as the value which minimizes $ \mathcal L(L,T,0)=\mathcal L [\bar\sigma_{L,T,h}(x_1),L,T,0]$.  
This allowed us to verify numerically the scaling exponents $\Gamma_{\bar L}$ reported in the first two columns of Table 1 for 
$T=0$  ($\triangle_J\!=\!-1$) and at the mean-field Curie temperature ($\triangle_J\!=\!-2$). Within the numerical accuracy we found complete agreement.   
The curves in Fig.~4a show some profiles $\bar\sigma_{L,T,0}(x_1)$ that are solutions of the variational equation $\delta \mathcal L[\sigma(x_1), \sigma_0(T)]/\delta (\sigma_0\, \sigma(x_1))\!=\!0$, 
but do not minimize -- in general -- the functional $\mathcal L (L,T,0)$ with respect to $L$  (for a striped structure in the thin DFIF). 
The black curve corresponds to the equilibrium $\bar L(T)$. The same profiles are plotted in Fig.~4b as a function of $y_1=x_1/L$, instead of $x_1$ (Fig.~4a).  
Indeed, from Fig.~4b, the shape of $\bar\sigma_{L,T,0}(y_1)$ turns out to be almost independent of $L$. 
This fact confirms that the scaling invariance proposed in Eq.~\eqref{scaling} is a plausible \textit{a priori} assumption.  
Fig.~4c shows the numerical $B(T)$ (black curve), calculated using the definition of Eq.~\eqref{modulus}   
within the mean-field approximation for the striped thin DFIF ($\delta\!\ll\!\bar L$).  
The displayed behavior virtually coincides with the scaling prediction (light curve), obtained dividing the calculated $\sigma_0^2(T)$ by the calculated $\bar L(T)$, in line with the corresponding scaling exponent in Table~1; 
the values of $\sigma_0$ and $\bar L$ at different temperatures are produced by the same calculation as $B$.  
The numerical instability observed at low $T$ in Fig.~4c is directly related to the fact that domain walls become sharper and sharper as $T\!\rightarrow \!0$; 
consequently, the continuum approach used to perform the present calculation becomes inappropriate (an analogous mean-field calculation within a discrete-lattice approach is reported in Ref.~\onlinecite{Ale}).   
Note how the anomalous growth of $B^{exp}(T)$ shown in Fig.~3, obtained from experiments on ultra-thin Fe films, is well-reproduced by our mean-field calculation.   
The same numerical check was performed for all the other cases reported in Table~1.
Fig.~4d shows the almost perfect coincidence between the numerical $B(T)$, computed 
directly from Eq.~\eqref{modulus}, and the scaling prediction, Eq.~\eqref{scaling_B}, 
for the thick DFIF ($\delta\!\gg\!\bar L$). 
In the inset, the same quantities are compared for the CFIF both in the thin ($\delta\!\ll\!\bar L$) and thick ($\delta\!\gg\!\bar L$) limit finding the same excellent agreement. 
Remarkably, $B(T)$ decreases smoothly with temperature and does not show any anomaly in all cases apart from the thin DFIF. 
\subsection{Monte-Carlo calculation for a 1d system} 
In order to provide a further numerical check of the scaling hypothesis we performed Monte-Carlo simulations with a standard Metropolis algorithm. 
The simulated system consist of $N$ Ising variables $\sigma_i=\pm 1$ aligned along a chain and ruled
by the discrete Hamiltonian of  Eq.~\eqref{discrete_lattice_Ham}. For computing the susceptibility we have used the variance in the magnetization which, in this particular case, is expressed as 
\begin{equation}
\chi(T) = \frac{1}{NT}\sum_{i,j}\langle\sigma_i\sigma_j\rangle\,,
\end{equation}
\begin{figure}
\includegraphics{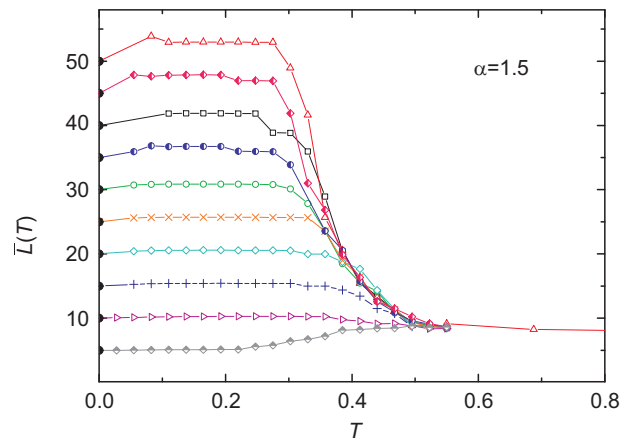}
\caption{Monte-Carlo simulations for a spin chain ($d=1$)
with $\alpha=1.5$  (color online). $\bar L(T)$, obtained form the position of the peak of 
the structure factor, as a function of $T$. Different symbols correspond to different initial configurations 
of a perfectly periodic pattern with period $\bar L(0)=5-50$ (solid dots on the $T=0$ axis).} 
\end{figure}
\noindent since the mean value of the magnetization is zero.
In spite of the lack of a magnetically ordered phase for $d\!=\!1$, we expect that the computed susceptibility
behaves like $\chi_A(T)$, Eq.~\eqref{Chi}, as long as the correlation length is significantly larger than the characteristic period
of modulation and the number of domain walls is nearly constant. 
Under these assumptions, a series of alternating domains is formed and the fluctuation of the  magnetization is mainly  due to the domain-wall displacements. 
This is equivalent to saying that domain-wall nucleation practically does not occur. 
By counting the number of domain walls during the susceptibility measurement we could confirm  
that the nucleation of new domain walls is a very rare event over the time spanned in our simulation and for the system sizes, $N$, we used.
For the specific case ($d\!=\!1$), the scaling behavior we want to check is  
\begin{equation}
\label{chi_scaling_1d}
\chi(T)\sim \bar L^{\alpha-1}\,.
\end{equation} 
Experimentally~\cite{Niculin}, the scaling behavior was tested at fixed $\alpha\!=\!3$, $d\!=\!2$ and for small $h$ by recording $\bar L$ -- while varying the temperature -- and checking that $\chi_A$ is a linear 
function of $1/\bar L(T)^{\triangle_g}$. 
\begin{figure}
\includegraphics{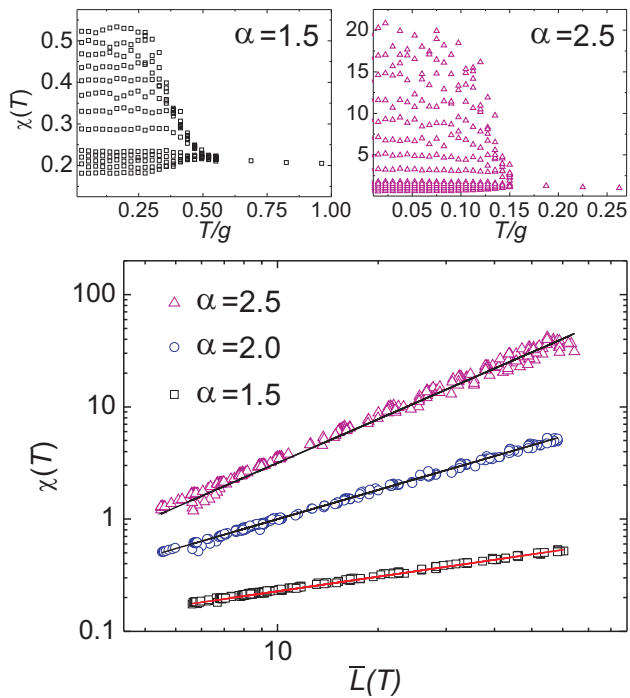}
\caption{Monte-Carlo simulations  (color online). Susceptibility, $\chi(T)$, for spin chains ($d=1$) with different $\alpha$.
Above, $\chi(T)$ is plotted versus $T$ for $\alpha=$1.5 (squares), 2.5 (triangles).
Below, $\chi(T)$ is plotted versus $\bar L(T)$ in a log-log scale  for  $\alpha=$1.5 (squares), 2 (circles) and 2.5 (triangles);
solid lines correspond to the scaling predictions $\chi(T)=C \left(\bar L(T)\right)^{\alpha-1}$, Eq.~\eqref{chi_scaling_1d},  
with $C$ adjusted parameter.} 
\end{figure}
In Monte-Carlo simulations, we chose a different route.  
$\bar L$ can be defined from the position of the highest peak in the structure factor~\cite{Cinti}. 
When the system is heated at a slow constant rate 
($\Delta T/\text{MCS} = 10^{-5}$, time measured in MC steps)  starting from different configurations at $T\!=\!0$, 
a set of \textit{metastable} states with a well-defined $\bar L(T)$ is produced. In this context, the meaning of   
$\bar L(T)$ is thus different from the rest of the paper: it represents a \textit{time} average for metastable states rather than an equilibrium period of modulation.    
In Fig.~5, $\bar L$ is reported for $\alpha=\!1.5$; different symbols correspond to different initial configurations.  
Periodic patterns of alternating domains with positive and negative magnetization were chosen as 
initial states (solid dots in Fig.~5 indicate the size of those domains, $\bar L(0)$). 
The different curves meet at $T/\lambda\simeq 0.5$, indicating that the period of modulation equilibrates only at higher temperatures. 
This behavior reminds some features of glassy transitions predicted to occur in analogous frustrated systems~\cite{Wol}. \\
The set of metastable states with different $\bar L$ generated at low temperature following this procedure can be used to verify the scaling law~\eqref{chi_scaling_1d}, see Fig.~6. The upper panel of Fig.~6 reports the susceptibility as a function of temperature, obtained for different initial configurations, for $\alpha=$1.5 and 2.5. As for  $\bar L(T)$ (Fig.~5), different initial configurations give different values of the 
susceptibility at a given temperature (for clearness, we don't indicate $\chi$ obtained for different initial configurations with different symbols). In the bottom panel of Fig.~6, the susceptibility data of the upper panel are plotted as a function of $\bar L(T)$, including values for   
$\alpha=$2. The log-log scale reveals that the power law behavior expected from scaling is indeed well-obeyed in the considered range of temperatures. The solid lines represent the scaling prediction, Eq.~\eqref{chi_scaling_1d}, with an adjusted multiplicative constant to fit the vertical shift.        
We remark that the scaling law just confirmed turns out to be fulfilled even when the system is locked in metastable states, where the value of $\bar L$ strongly deviates from its equilibrium average.   
\section{Stability against fluctuations\label{sec_VII}}
The variational approach introduced in the previous Sections~\ref{sec_IV} and~\ref{sec_VI}  
produces a non-vanishing -- typically highly symmetric -- equilibrium profile $\bar\sigma_{\bar L,T,0}(\vec x)$ of the order parameter at any $T\!<\!T_C$, such as the mono-dimensional order represented by a striped structure. This result implies the existence of positional long-range order. Yet one-dimensional modulated order described by $\bar\sigma_{\bar L,T,0}(x_1)$  
-- such as the one represented by stripes -- is strictly speaking forbidden in the thermodynamic limit by the Landau-Peierls instability~\cite{Poki}. 
It is therefore reasonable to ask whether those fluctuations which destroy the positional mono-dimensional order at finite temperatures -- not included in the variational approach -- might also affect the
results following from the scaling hypothesis, Eq.~\eqref{scaling}. 
In the following, we test the stability of our scaling hypothesis against small (elastic) deviations from the strict mono-dimensional order in a two-dimensional system (in-plane coordinates $x_1,x_2$). 
Suppose that the variational mean-field equations have found an equilibrium stripe profile  $\bar\sigma_{\bar L,T,0}(x_1)$ 
corresponding to a certain equilibrium stripe width $\bar L$ (to simplify the notation, here we drop the dependence on all the parameters but $x_1$ in $\bar\sigma$). 
Let us slightly modify this one-dimensional profile $\bar\sigma(x_1)$  
along the direction of modulation, $x_1$, and along the direction perpendicular to it, $x_2$, into 
$\bar\sigma\!\left(x_1\!-\!\epsilon_{k_1,k_2}\cos(k_1 x_1)\cos(k_2 x_2)\right)$. 
This replacement establishes an elementary excitation displacing domain walls along the $x_1$-coordinate and bending them along the $x_2$-direction and accounts for the experimental fact that, e.g. stripes are never perfectly straight in real systems at finite temperatures~\cite{Koker}. 
These are the excitations that prevent one-dimensional \textit{positional}  long-range order in $d\!=\!2$ and ultimately invalidate the mean-field approach in this respect~\cite{Garel,Poki,Muratov}. 
One can show that the perturbed functional $\ell \left[\bar \sigma\left(x_1\!-\!\epsilon_{k_1,k_2}\cos(k_1 x_1)\cos(k_2 y_2)\right) ,\bar L,T,0\right]$, 
expanded up to the order $\epsilon_{k_1,k_2}^2$, writes as the sum of the original one $\ell [\bar \sigma(x_1),\bar L, T,0]$ plus corrections arising from the short-ranged and long-ranged interaction
\begin{equation}
\begin{split}
&\Delta\ell =\epsilon_{k_1,k_2}^2\frac{J}{8}\delta\Lambda \sigma_0^2 (k_1^2 + k_2^2) \int (\partial_{x_1} \bar \sigma(x_1))^2  dx_1\\ 
&+\epsilon_{k_1,k_2}^2\frac{\pi}{4}\delta^2 \lambda \Lambda \sigma_0^2 \iint \bar \sigma(x_1)\bar\sigma(x_1') \\
&\times\int q^2 e^{iq(x_1-x_1')}  
\Bigl[G_\alpha\left(\sqrt{(q+k_1)^2\!+\!k_2^2}\right) \\
& + G_\alpha\left(\sqrt{(q-k_1)^2\!+\!k_2^2}\right)\! -\!2G_\alpha(q)\Bigr] dq\, dx_1\, dx_1'\,. 
\label{fluc}
\end{split}
\end{equation}
The fluctuation correction, Eq.~\eqref{fluc}, can be evaluated, up to the order $k_1^nk_2^m$, $n+m=4$ using the equipartition theorem~\cite{Poki},
\begin{equation}
\langle \epsilon_{k_1,k_2}^2\rangle\sim \frac{T}{\Lambda^2\, g \sigma_0^2\,\left[F_1(\alpha)\,q^{\alpha-2} k_1^2+F_2(\alpha)\,q^{\alpha-4} k_2^4\right]}\,,
\label{fluct_spectrum}
\end{equation}
($F_1(\alpha)$ and $F_2(\alpha)$ being some constants) and summing over $(k_1,k_2)$. 
Eq.~\eqref{fluct_spectrum} describes the spectrum of fluctuations for $q\sim 2\pi/\bar L$.  
As $\bar \sigma(x_1)$ is periodic with period $\bar L$ by definition, the integral $\int \cdots dq$ is non-vanishing only for $q\ge 2\pi/\bar L$, thus one can essentially identify 
$q$ with any Fourier component of the profile $\bar \sigma(x_1)$.  
The sum over $(k_1,k_2)$ takes into account all fluctuations with wave length larger than the natural cut-off $\bar L$, namely $(k_1,k_2)<1/\bar L$. 
After treating the fluctuation corrections with this procedure,
one recovers an effective mono-dimensional functional in which the 
contribution of both the short-ranged and the long-ranged interaction have, to leading order, the same scaling behavior as the functional~\eqref{functionals}, 
albeit with modified coupling constants 
\begin{eqnarray}
\begin{cases}
J \mapsto J\times\left[1+ {\mathcal O}\left(\frac{T}{J}\right)\right]\\
g \mapsto g\times\left[1+ {\mathcal O}\left(\frac{T}{J}\right)\right] \,.
\end{cases}
\end{eqnarray} 
This allows us to conclude that the scaling results are not affected by the presence of this kind of fluctuations at finite temperature. 
In particular, we expect $\triangle_g$ and the scaling dimension with $\bar L$ of any quantity not to be changed by such small perturbations. \\ 
A further argument validating the concept of a stable equilibrium modulation length is a kind of Ginzburg-Landau criterion measuring $\langle (L\!-\!\bar L)^2 \rangle/\bar L^2$. 
Using the scaling functional of Eq.~\eqref{functionals} in the vicinity of the putative equilibrium stripe width   
\begin{equation}
\ell (L,T) = \ell(\bar L,T) + (L-\bar L)^2\frac{\partial^2 \ell (L,T,0)}{\partial L^2}\Bigg|_{\bar L} \,,
\end{equation}
the scaling property 
\begin{equation}
\frac{\partial^2 \ell (L,T)}{\partial L^2}\Bigg|_{\bar L} \sim g \sigma_0^2(T)\, {\bar L}^{\triangle_g-2}
\end{equation}
and the equipartition theorem, we obtain 
\begin{equation}
\frac{\langle (L-\bar L)^2 \rangle}{\bar L^2}\!\sim \!\frac{T \, \bar L^{-\triangle_g}}{V_d\, g\,\sigma_0^2(T)}\,.
\end{equation}
The latter expression diverges
only in the vicinity of $T_C$, because of the vanishing $\sigma_0(T)$ (remember that $\bar L$ obtained from  the variational approach is finite for $T\!\leq\! T_C$~\cite{Ale}). 
Accordingly, $\bar L$ remains a well-defined quantity almost over the entire temperature range where $\sigma_0(T)$ is finite. 
In the vicinity of $T_C$, for $\alpha\!=\!3$ and $d\!=\!2$, the mean-field approximation~\cite{Oliver2} predicts $\bar L(T) = \bar L(T_C) + L_1\left(1-T/T_C\right)^2$ and 
$\sigma_0(T)\sim\left(1-T/T_C\right)^\beta$ with $\beta=1/2$.       
In ultra-thin Fe films on Cu(001), when approaching the paramagnetic phase from lower temperatures, $\bar L(T)$ is found to obey the mean-field behavior reported above;  
remarkably, in the same temperature range $\sigma_0(T)$ does \textit{not} show the mean-field critical exponent, i.e. $\beta\!\ne\!1/2$ (see Fig.~2c in Ref.~\onlinecite{Niculin}).  \\
In summary, the Landau-Peierls instability destroys long-range positional stripe order but the loss of order proceeds in such a way that the modulation length 
and its relationship with other observables remain meaningful also at finite temperatures.
This last statement is supported by  Monte-Carlo results for spin chains presented in Section~\ref{sec_VI} and previously~\cite{Cinti}.
\section{Conclusions\label{sec_VIII}}
We have presented a scaling procedure that is generally applicable to any pattern-forming system. This procedure 
allows rewriting the Hamiltonian~\eqref{functional} as a functional of scaling-invariant scalar fields with coupling constants rescaled by 
appropriate powers of some emerging characteristic spatial scale $L$.  
In virtue of this scaling hypothesis, the temperature dependence of $\bar L$ (expectation value of $L$) propagates in a non-trivial, but predictable way to any observable. 
The consequences of this scaling relations can be dramatic because $\bar L$ may vary by several orders of magnitude, as in the thin DFIF.  
Our theoretical predictions are confirmed by a number of experimental and computational results (see Section~\ref{sec_IV} and~\ref{sec_VI}).\\ 
One implication of the scaling  analysis is the existence of a class of systems, defined by the inequality
\begin{equation}
d<\alpha\leq d-\triangle_J \,,
\label{ISB}
\end{equation}   
showing some anomalous physical behaviors. 
The thin DFIF and an equivalent model of Coulomb-frustrated phase separation~\cite{Kivelson}  belong to this class. 
In the range of values of $\alpha$ defined by the inequality~\eqref{ISB} two remarkable effects may occur. \\
First, as the modulation length decreases with increasing temperature, the compression modulus $B$ may \textit{increase} with temperature according to Eq.~\eqref{scaling_B}.  
The class of matter defined by the inequality~\eqref{ISB} exhibits a state of potentially anomalous compression modulus in which, e.g., it becomes more difficult for the system to accommodate disorder involving deviations from the equilibrium modulation length when the temperature is increased; the system becomes {\it less} responsive to fluctuations at higher temperatures. 
This behavior is at odds with what normally happens.  
In fact, traditional elastic constants 
are, in the harmonic approximation, temperature independent and the small correction due to anharmonic contributions typically soften them with increasing temperature~\cite{x}. 
For the thin DFIF, we tentatively associated this hardening of the compression modulus, $B$, 
with the re-entrance of stripe order during pattern transformations. 
Although we do not know how $B$ relates to the patterns, we suggest that our results indicate the need of refining the various scenarios for topological phase 
transitions~\cite{KTNHY} to take into account the role of anomalous behavior of elastic  constants. \\
Second, the scaling hypothesis, Eq.~\eqref{scaling}, predicts that lines marking the transition between two different phases on the $h\!-\!T$ plane have the shape of a ``funnel'' 
rather than of a ``dome'' (like in traditional phase diagrams).  This anomalous topology has an important consequence.  
The vast majority of phase transitions in condensed matter, particle physics and cosmology proceeds 
by a mechanism known as Symmetry Breaking (SB)~\cite{Landau,Nambu,Weinberg,Rivers}. 
As pointed out in Section~\ref{sec_IV} and sketched in Fig.~2, only SB transitions are compatible with a dome-like phase diagram~\cite{Seul,Garel,Gennes,Fogler,St,Kivelson,Muratov,Poki}: 
The phase outside the dome (high $T$ or high $h$) has full rotational and translational symmetry, while the region enclosed by the dome boundaries corresponds to a phase with less symmetry elements. 
Thus, the degree of symmetry of two different phases involved in a transition can only be reduced by lowering the temperature.  
This rule, governing the ordinary scenario of thermodynamic phase transitions, can be violated when a phase diagram displays a funnel shape and 
ISB transformations may also occur~\cite{Landau,Weinberg} (see Fig.~2b). 
Such transitions, though rarely, have been indeed observed in condensed matter physics~\cite{Schnerb,Stillinger,Ruocco,Bec,Oliver1}.
The experimental phase diagram of ultra-thin Fe films on Cu(001), a model system for thin DFIF ($\alpha\!=\!3$ and $d\!=\!2$), shows the funnel shape predicted by our scaling hypothesis~\cite{Niculin}. 
Along some specific paths on the $h\!-\!T$ plane a SB transition from a uniform pattern to a striped phase -- which breaks both 
the full rotational and translational symmetry --  is observed upon lowering the temperature; 
when the temperature is further lowered, the following ISB processes are also encountered: 
\begin{itemize} 
\item a stripe-to-bubble phase transition, which restores the rotational symmetry 
\item a bubble-to-uniform phase transition, which restores the full translational symmetry. 
\end{itemize}
These experimental observations strongly support our scaling hypothesis. 

In conclusion, our scaling hypothesis predicts that the functional in Eq.~\eqref{functional} is compatible with both SB and ISB processes. 
Note that the functional discussed in this paper is fundamentally different from the one described by recent models introduced to explain ISB anomalies in condensed matter physics~\cite{Schnerb,Stillinger}. 
In particle physics and cosmology, a model of ISB based on a multi-component scalar field has been proposed in the context of unresolved fundamental problems such as the CP-violation during baryogenesis, 
the inflation of the universe or the formation of cosmological defects~\cite{Weinberg,Ramos,Viet,Bajc,Sen,Green}. 
Here, instead, we always deal with a single component scalar field.  
Another remarkable difference with respect to current models of ISB lies on the fact that whether the Hamiltonian in Eq.~\eqref{discrete_lattice_Ham} can produce SB or ISB processes just depends on $d$ and $\alpha$, not on the particular choice of the coupling constants or on some \textit{ad hoc} assumptions. 
We can state therefore that, when competing interactions are involved, SB and ISB must be treated on equal footing: there is, in principle, no \textit{a priori} predominance of SB processes.
\begin{acknowledgments}
We acknowledge the financial support of ETH Zurich and the Swiss National Science Foundation.
\end{acknowledgments}

\vspace{0.5cm}
\appendix
\section{Fourier transform of the kernel $G_\alpha$\label{Appendix_A}}
Integrals of the type
\begin{equation}
\int_{{\cal R}^d} \frac{1}{|\vec r|^\alpha}\phi(\vec r)dV
\end{equation}
are well-defined within the theory of distributions~\cite{Dono1,Dono2}. The set of  test functions describing the scalar field $\phi(\vec r)$ is supposed to be well-behaved for $|\vec r|\rightarrow \infty $. The distribution $1/|\vec r|^\alpha$ is locally integrable for $\alpha<d$. It can be continued analytically to $\alpha\geq d$ provided the test functions (and a sufficiently large number of their derivatives) are taken to be vanishing at $\vec r =0$~\cite{Dono1}. This can be performed, e.g., by appropriate subtraction~\cite{Dono1}. This corresponds to eliminating the (unphysical) singularity of the kernel $G_\alpha(|\vec x-\vec x'|)$ at $\vec x=\vec x'$. Notice that the following results do not depend on how exactly the test functions are made to be vanishing (introduction of a cut-off, range of the cut-off~\cite{Singer}, etc.).
In the range $\frac{d-1}{2}\!<\!\alpha\!<\!d$ there exists an analytic Fourier transform
\begin{eqnarray}
\frac{1}{(2\pi)^d}\int  \frac{1}{|\vec r|^\alpha} \,e^{i\vec k\cdot\vec r} \,dV  
\end{eqnarray}
which amounts to~\cite{Dono1}
\begin{eqnarray}
G_{\alpha}(\vec k)= \frac{1}{
2^\alpha \pi^{d/2}}\frac{\Gamma(\frac{d-\alpha }{2})}{\Gamma(\frac{\alpha }{2})} | \vec k| ^{\alpha -d}\,.
\label{FT1}
\end{eqnarray}
The question arises about the Fourier transform of the (subtracted) distribution outside the range of validity of Eq.~\eqref{FT1}. As shown, e.g., in Ref.~\onlinecite{Dono1,Dono2}, one can continue the right-hand side of Eq.~\eqref{FT1} to any value of $\alpha$ to obtain the Fourier transform of the subtracted distribution provided the analytical continuation is well-defined. However, the right-hand side of Eq.~\eqref{FT1} is ill-defined for $\alpha = -2l$ and
$\alpha =d+2l$, $l\in {\cal N}_0$. For these values the Gamma functions entering Eq.~\eqref{FT1} have poles. Thus, for these special values of $\alpha$ -- some being physically relevant, as pointed out in the paper -- the Fourier transform must be evaluated \textit{ad hoc}.
One can immediately recognize that~\cite{Dono2}
\begin{equation}
G_{-2l}(\vec k)= (-\triangle )^l\delta(\vec k)  \,.
\label{FT2}
\end{equation}
Both $G_\alpha $ and $G_{-2l}$ are homogeneous distributions.\\
The case $\alpha\!=\!d\!+\!2l$ is more cumbersome and the outcome depends on the exact definition~\cite{Dono1,Dono2} of the distribution $1/|\vec r|^\alpha$.
We compute $G_{d+2l}(\vec k)$  
in spherical coordinates ($dV\doteq r^{d-1}dr\, d\Omega$, $d\Omega$ being the infinitesimal surface element of a unit sphere in $d\geq 2$-dimensions).
Performing the integral over the angular variables leads to
\begin{equation}
G_{d+2l}(k) = \frac{\Gamma(\frac{2l+1}{2})}{2^{d-1}\pi^{\frac{d+1}{2}}\Gamma(\frac{2l+d}{2})}\, \int \frac{\cos(k r)}{r^{2l+1}} \, dr\,.
\end{equation}
We treat the radial integral, extending from $0$ to $\infty$, as a distribution and the integral
\begin{equation}
\int \left[\int \frac{\cos(kr)}{r^{2l+1}} dr \right]\phi(k)\, dk
\end{equation}
is evaluated through repeated integration by parts. The parts integrated out are forced to vanish by assuming that the test functions $\phi(r)$ and a sufficient number of their derivatives vanish at $r=0$, so that we can write
\begin{eqnarray}
\int  \frac{\cos(kr)}{r^{2l+1}} \,dr &=& \frac{(-1)^l}{(2l)!}\, k^{2l}\int \frac{\cos kr}{r} \,dr\nonumber\\
&=& \frac{(-1)^l}{(2l)!}\,k^{2l}\int \ln(r) \sin (kr) \, dr \nonumber\\
&=& \frac{(-1)^l}{(2l)!}\, k^{2l}\, \Bigl(- \ln k\Bigr) \,.
\end{eqnarray}
Summarizing, we get
\begin{equation}
G_{d+2l}(\vec k)= \frac{2(-1)^l}{\pi^{d/2}2^{d+2l}\Gamma(\frac{2l+d}{2})l!}|\vec k|^{2l}\Bigl(-\ln|\vec k|\Bigr)
\label{FTspecial}
\end{equation}
which can be continued to $d\geq 1$. This distribution contains a logarithmic part and is therefore not exactly homogeneous. Notice that some authors~\cite{Dono1,Dono2} add an extra term $\propto |\vec k|^{2l}$ in Eq.~\eqref{FTspecial}, depending on their definition of the distribution $1/|\vec r|^{d+2l}$. By eliminating the parts integrated out we do not obtain such a term. 
\vspace{0.5cm}
\section{Stability of the modulated phase\label{Appendix_B}}
For the stability analysis we need to compute the compression modulus $B(T)$ 
rigorously, i.e. beyond the variational approach. A rigorous thermodynamic 
treatment of the functional $\ell(\sigma(\vec y),L,T,h)$ defines the functional 
$\ell(L,T,h)$ by means of the generalized Hellmann-Feynman theorem~\cite{theorem}, 
which equates the derivative of the $\ell(L,T,h)$ with respect to the parameter $L$ 
with the functional average over all configurations $\sigma(\vec y)$ of the 
derivative of $\ell[\sigma(\vec y),T,L,h]$ with respect to $L$:
\begin{eqnarray}
\frac{\partial \ell(L,T)}{\partial L}&\doteq&\left\langle \frac{\partial}{\partial L}\ell[(\sigma(\vec y),L,T,h]\right\rangle_{\sigma(\vec y)}\\
\left\langle\cdots\right\rangle_{\sigma(\vec y)}&\doteq&
\frac{\int D\sigma(\vec y)\cdots e^{-\beta V_d \ell[(\sigma(\vec y),L,T,h]}}{\int D\sigma(\vec y) e^{-\beta V_d \ell[(\sigma(\vec y),L,T,h]}}\,.
\end{eqnarray} 
The characteristic length $\bar L$ is defined as the solution of the equation $\partial \ell(L,T,h)/\partial L\!=\!0$.
The compression modulus $B(T)$, see Eq.~\eqref{modulus}, is computed accordingly:
\begin{eqnarray}
B(T)&=&\bar L^2\!\left\langle\! \frac{\partial^2}{\partial L^2}\ell[\sigma(\vec y),L,T,h] \right \rangle_{\!\!\!\sigma(\vec y)}\!\Bigg |_{\bar L} \nonumber \\
\!&-&\!\frac{1}{T}\,\bar L^2\!\left\langle\!\left(\!\frac{\partial}{\partial L}\ell[\sigma(\vec y),L,T,h]\!\right)^{\!\!2} \right \rangle_{\!\!\!\sigma(\vec y)}\!\Bigg |_{\bar L}\,.
\label{B}
\end{eqnarray}
The scaling analysis, applied  to Eq.~\eqref{B}, leads to 
\begin{eqnarray}
B(T)&\sim&\bar L^2\, J\, \Bigl[{\cal O}\Bigl((g/J)^{\frac{2-d+\alpha}{d-\alpha-\triangle_J}+1}\Bigr)\nonumber\\
&-& \frac{J}{T}\,{\cal O}\Bigl((g/J)^{\frac{2-2d+2\alpha}{d-\alpha-\triangle_J}+2}\Bigr)\Bigr] \,.
\end{eqnarray}
The positivity of $B$ is a necessary condition for the stability of the modulated state. In the limit $g/J\!\rightarrow\!0$, positivity of $B$ requires the exponent of the first term to be smaller than the exponent of the second (always negative) term, i.e. Eq.~\eqref{stability}, $\alpha\!-\!d\! < \!-\!\triangle_J$.


\begin{thebibliography}{100}
\bibitem{cc}
$J$ is of the order of $T_C$, the Curie temperature. The coupling constant $g$ amounts, for a system with atomic spin $S$ and subject to the dipolar interaction, to $\mu_0(g\mu_BS)^2\cdot (4\pi a^3)^{-1}\doteq \Omega$ and to $q^2\cdot (4\pi \epsilon_0 a)^{-1}\doteq Q$ for charges of strength $q$ interacting via the Coulomb interaction. For a spin system, $h\!=\!\mu_0(g\mu_B S)\!\cdot\!H$, where $H$ is an externally applied magnetic field.
\end{thebibliography}
\end{document}